\documentclass[a4paper,11pt]{article}
\pdfoutput=1 

\usepackage{jinstpub} 

\usepackage{graphicx}
\usepackage{float}
\usepackage{flafter}
\usepackage{textcomp}
\usepackage{lipsum}                                                 
\usepackage{mwe}                                                             

\title{iDREAM: Industrial Detector of REactor Antineutrinos for Monitoring at Kalinin nuclear power plant}

\author[a]{A.~Abramov,}
\author[b]{A.~Chepurnov,}
\author[a,c]{A.~Etenko,}
\author[b,d]{M.~Gromov,}
\author[a]{A.~Konstantinov,}
\author[a]{D.~Kuznetsov,}
\author[a,c,1]{E.~Litvinovich,\note{Corresponding author.}}
\author[e]{G.~Lukyanchenko,}
\author[a,c]{I.~Machulin,}
\author[a]{A.~Murchenko,}
\author[a]{A.~Nemeryuk,}
\author[a]{R.~Nugmanov,}
\author[a]{B.~Obinyakov,}
\author[a]{A.~Oralbaev,}
\author[a]{A.~Rastimeshin,}
\author[a,c]{M.~Skorokhvatov}
\author[a]{and S.~Sukhotin}

\affiliation[a]{National Research Centre "Kurchatov Institute", Moscow 123182, Russia}
\affiliation[b]{Skobeltsyn Institute for Nuclear Physics, Lomonosov Moscow State University, Moscow 119234, Russia}
\affiliation[c]{National Research Nuclear University "MEPhI" (Moscow Engineering Physics Institute), Moscow 115409, Russia}
\affiliation[d]{Joint Institute for Nuclear Research, Dubna 141980, Russia}
\affiliation[e]{Lomonosov Moscow State University Faculty of Physics, Moscow 119991, Russia}

\emailAdd{Litvinovich\_EA@nrcki.ru}

\abstract{The paper is devoted to the description of the iDREAM detector and its systems. iDREAM is a prototype detector designed to demonstrate the feasibility of antineutrino detectors for remote reactor monitoring and safeguard purposes. Antineutrinos are detected with a 1 ton liquid scintillator via inverse beta decay on protons. In order to suppress cosmic muons, gamma and neutron background, the detector is housed in a dedicated shielding. The detector is installed at the Kalinin nuclear power plant (Russia), 20 m from the 3 GW$_{th}$ reactor core.}


\begin{document}
\maketitle
\flushbottom

\section{Introduction}
Neutrino studies at nuclear reactors have a long history, starting with the experiment by Cowan and Reines, which confirmed the existence of the neutrino~\cite{Cowan+_1956}. In nuclear reactors, electron antineutrinos (referred hereinafter simply as “neutrinos” for brevity) originate from beta-decay chains of fission products. On average, about 6~neutrinos per fission are produced, thus providing a large neutrino flux of the order of 10$^{20}$ $\nu$/s per GW of thermal power (GW$_{th}$). The spectrum of reactor neutrinos goes up to about 8~MeV.

Even with such an intense source as nuclear reactors, neutrino interaction rates are still very small, so one has to use large, uniquely designed, detectors and collect statistics for long time periods. Because of this, early neutrino experiments were focused on fundamental studies exclusively.

The idea to utilize neutrinos as a tool for nuclear reactor monitoring was first formulated in the late 1970s by L.~Mikaelyan ~\cite{Mikaelyan_1978} and then elaborated in ref.~\cite{Borovoi+_1978}, giving rise to the field known today as ''applied antineutrino physics''. More specifically, two ideas were pointed out:

\begin{itemize}
  \item The neutrino flux (and, hence, the rate of detected events) is directly proportional to the energy output of the reactor. Therefore, by measuring the neutrino event rate, one can determine the reactor thermal power.
  \item Each of the fissile isotopes has a unique neutrino spectrum associated with it. As the fuel burns up, the contribution of each isotope to the overall reactor spectrum (and to the total number of neutrinos) changes. This allows one to observe the evolution of fissile inventory (in particular, the accumulation of plutonium) by measuring the neutrino spectrum and/or total event rate.
\end{itemize}

Following refs.~\cite{Mikaelyan_1978, Borovoi+_1978}, a number of proof-of-principle studies were performed by Mikaelyan group at Rovno nuclear power plant (NPP) in the 80s and 90s (see ref.~\cite{Rovno_experiments} and references therein). In particular, it was demonstrated that one can determine the reactor power and fuel burnup remotely with neutrino detection. Later on, in ref.~\cite{Vyrodov+_1995}, the power levels of Rovno and Bugey reactors were directly compared by means of neutrino detection.

Since the early 2000s, the interest to reactor neutrino applications has been growing: demonstration experiments were made (in particular, SONGS1 \cite{Bowden+_2007}), new detection techniques were developed and more detailed case studies were done (for a review, see, e.g., refs.~\cite{Bernstein+_2020, Stewart_2019}). Altogether, these studies have demonstrated the feasibility of reactor monitoring by neutrinos, leading to an idea of industrial detectors for~NPPs. The necessity of further R\&D in this field was acknowledged by IAEA~\cite{IAEA_2008}.

The Industrial Detector of REactor Antineutrinos for Monitoring (iDREAM) is specifically intended as a prototype of a commercial detector that can be placed in NPPs. The concept was to use a simple design and use well-established technologies, thus offering easy manufacturing and maintenance for future detectors of this type. In 2021, the detector has been installed and commissioned at Kalinin~NPP (Russia) at 20~m from the 3~GW$_{th}$ reactor core (third unit).

In this paper, we describe the design and features of the iDREAM detector and its systems. The paper is organized as follows. In Section~2, we give a general description of the detector. Section~3 is devoted to detector shielding. The details on the scintillator production and the detector loading are given in Section~4. In Section~5 we give the description of the detector subsystems. Section~6 introduces the results of the detector calibration. We conclude the paper in Section~7.

\section{General description of the detector}
In iDREAM, neutrino interactions with a 1~m$^{3}$ gadolinium-doped liquid scintillator are detected by inverse beta decay (IBD)

\begin{equation} 
\widetilde{\nu_e} + p \rightarrow e^+ + n
\label{eq1}
\end{equation} 
with a threshold $E_{\nu}$ = 1.806 MeV. The process has the largest cross section, compared to other quasielastic neutrino-nucleus processes and neutrino-electron elastic scattering. IBD also allows to employ the time-coincidence method and suppress the backgrounds: the positron quickly annihilates, providing a prompt signal with $E_\mathrm{prompt}$ = $E_{\nu}$ - 0.784~MeV, while the neutron is thermalized and captured in the medium, giving rise to a delayed signal. 

\begin{figure}[htbp]
\centering
\includegraphics[scale=1.2]{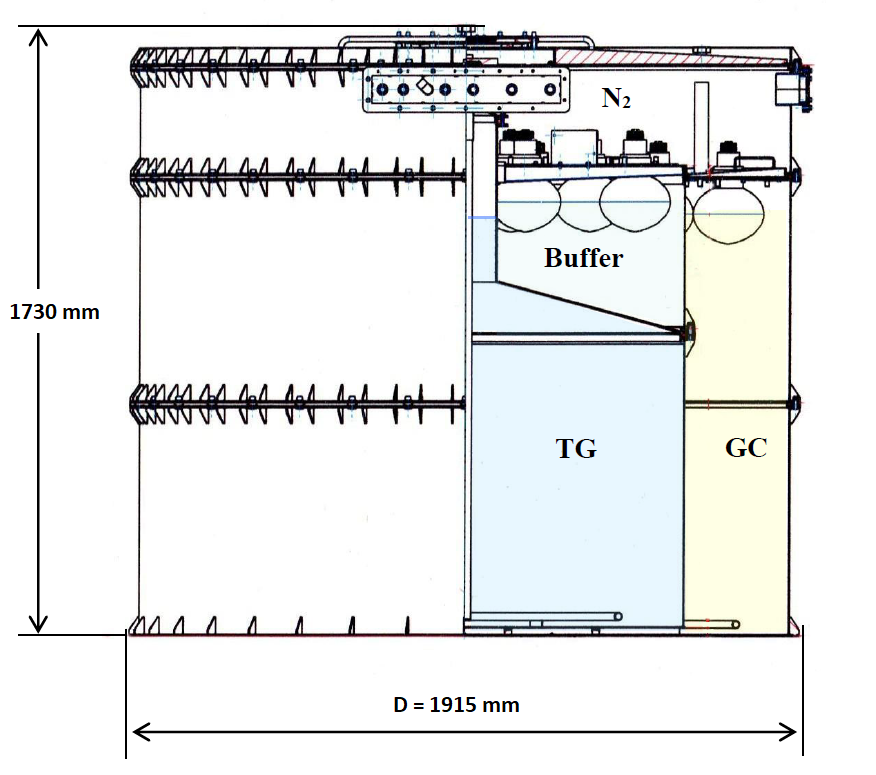} 
\caption{\label{fig1} Schematic drawing of the iDREAM detector.}
\end{figure}

A schematic drawing of the iDREAM detector is shown in figure \ref{fig1}. The detector consists of two concentric tanks made of 2~mm thick stainless steel. The inner tank has a diameter of 1254~mm and a height of 1320~mm, while the outer is 1858~mm in diameter and 1620~mm in height. The tanks are covered with a single pressure-sealed stainless steel cap (SSC).

The Inner Tank (IT) is rigidly fixed to the bottom of the outer tank. At a height of 835~mm its volume is vertically divided into two parts by means of a convex transparent acrylic membrane with a vertical tube along {\it z} axis. The tube is 180~mm in diameter and 470~mm in height. The bottom part of the IT, i.e. the 1.1~m$^{3}$ volume restricted by its walls and the membrane, is filled with Gd-LS based on Linear Alkylbenzene (LAB) and acts as a target (TG) for antineutrinos. The top part, 0.4~m$^{3}$ in volume above the membrane, is filled with pure LAB and acts as a buffer, shielding the target from the radioactive impurities coming from 16 photomultiplier tubes (PMT), which are installed on top of the IT in an inner stainless steel cap. The PMT photocathodes are immersed in a buffer LAB, thus PMTs view the target through a transparent buffer and a membrane. In order to increase the light collection, the inner walls and the bottom of the IT are covered with Lumirror foil, tightly pressed against steel surfaces by means of a spring-loaded steel grid.

The membrane tube extends above the IT. At the temperature variation within $(20\pm5)$ $^{\circ}$C the Gd-LS level neither exceeds the tube nor falls below its base. Nonetheless, an overflow reservoir, 10 liters in volume, is connected to the TG through a flexible tube. Two vertical tubes, 36~mm in diameter, are installed inside the membrane tube. The tubes go down to the floor of the IT. The one made of stainless steel, with the blind end, is used for deployment of the calibration sources. The other is made of acrylic and houses the Gd-LS temperature and level sensors.

The Outer Tank (OT) consists of three sections, sealed together by interflanged viton ring seals. The annular volume between the walls of the IT and OT, 1220~mm in height, is filled with 1.7~m$^{3}$ of the scintillator without gadolinium and acts as gamma-catcher (GC). Functionally, the GC aims at actively shielding the detector from cosmic muons, as well as increasing its efficiency through the detection of the IBD products that escape the TG. The GC is viewed by 12 PMTs, installed in pairs in six annular segments of the cap. Between the PMTs in each segment, six vertical calibration tubes, similar to that in the TG, are installed. The inner walls of the GC are covered by Lumirror reflective foil, while its floor is covered by Teflon sheets.

In the center of the outer cap covering both the IT and OT, there is a hole 250~mm in diameter, with a flange raised by 50~mm relative to the outer flange. The hole is closed with a ''floating'' cap sealed with a viton gasket. The ''floating'' cap contains 3 holes reproducing the geometry of the membrane cap holes. Each hole is covered with a sealed removable dummy disk. The ''floating'' cap aligns the vertical axes of the holes to the vertical axes of the corresponding holes in the membrane cap. 

\begin{figure}[htbp]
\centering 
\includegraphics[scale=0.6]{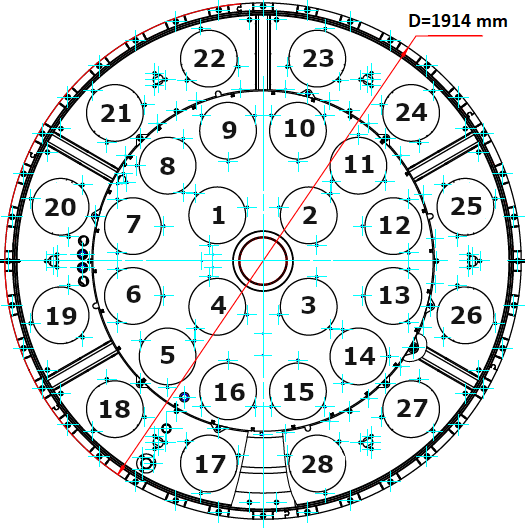} 
\caption{\label{fig2} PMTs location in the iDREAM detector: sixteen PMTs (1-16) are installed in the IT and view the TG, while twelve (17-28) are installed in the OT and view the GC.}
\end{figure}

The PMTs used are Hamamatsu R5912. In total, 28 PMTs are installed as shown in schematic figure \ref{fig2}: 16 PMTs in the IT and 12 PMTs in the OT.

Two rectangular hermetic windows are made in the upper part of the OT at a central angle of 90$^{\circ}$ to diffuse the internal communications of the detector. One contains a fluid distribution header (FDH), which is used to fill and empty the detector, purge the scintillator with nitrogen, and fill the space under the SSC with nitrogen. The other houses the assembly of the connectors for the PMTs high voltage supply and the reception of signals from the detector sensors.

The scintillator leak protection is ensured by a 700~mm high stainless steel emergency tray mounted under the detector.

\section{Detector shielding}
The detector is located at ground level beneath the reactor, which should provide $\sim$50 m.w.e. protection from cosmic muons. Meanwhile, still the cosmic muon flux can give a non-negligible impact on the background of the detector. At the same time, the gamma and neutron backgrounds at Kalinin NPP are expected to be higher than in a typical building ~\cite{DANSS}. This imposes strict requirements for the passive shielding.

\subsection{Passive shielding}
The passive shielding of the iDREAM detector has been designed with a particular emphasis on suppressing the ambient neutron and gamma radioactive backgrounds. Both contribute to random coincidences of events that may mimic the IBD signature.

The shielding surrounds the detector on every side. It is rigidly mounted on the modular support structure, which forms an inner cavity wherein the detector is placed. The base of the support structure is a rectangular 3800 $\times$ 4000 mm skid platform made of stainless steel U channels. At the center of the platform, 81 blocks of cast iron, 280 $\times$ 280 $\times$ 140~mm each, forming a square with a side of 2550 mm, are placed. The detector is installed in the center of the platform, on a stainless steel square support with a side of 2000 mm and a height of 250 mm. The space inside the support, i.e. between the cast-iron blocks and the detector floor, is filled with a layer of 80 mm thick pure polyethylene, and a layer of 100 mm thick borated polyethylene. These layers of cast iron and polyethylene make up the bottom part of the detector shielding.

\begin{figure}[htbp]
\centering 
\includegraphics[scale=0.6]{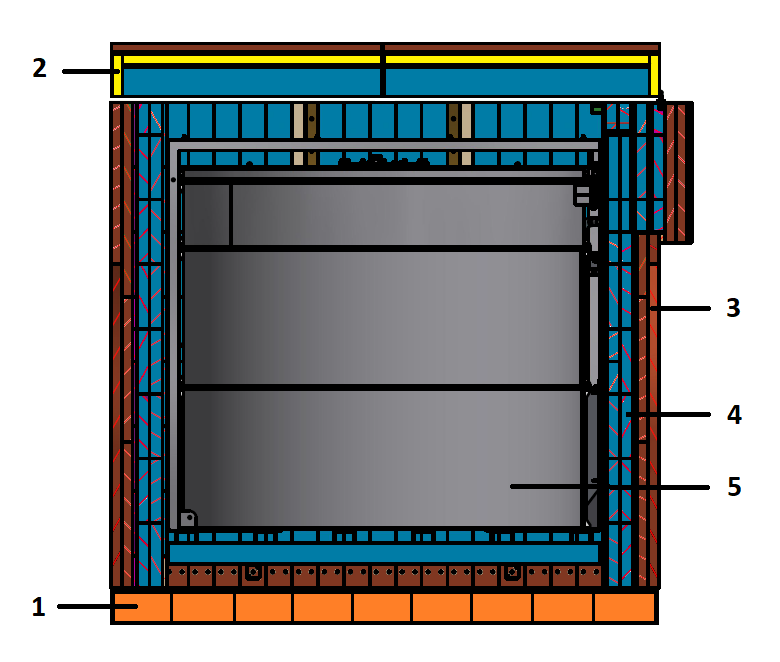} 
\caption{\label{fig3}Schematic drawing of the detector inside the shielding: 1 (orange) -- cast iron, 2 (yellow) -- lead, 3 (brown) -- pure polyethylene, 4 (blue) -- borated polyethylene, 5 -- detector.}
\end{figure}

Along the perimeter of the detector, the combined passive shield screens fixed on the pad of the support structure are installed. The screens consist of four layers: two inner ones, each 80~mm thick, constructed from borated polyethylene "NEUTRONSTOP" C-type bricks, and two outer ones, each 50~mm thick, which are assembled from pure polyethylene plates. The installation of additional, outermost shielding made of 14 cm thick cast-iron blocks is intended but not yet implemented. To access the FDH and the connectors assembly, two removable windows are made in the corresponding parts of the shield screens. The schematic drawing of the detector shielding is shown in figure \ref{fig3}.

The upper part of the passive shielding is a square 2550~mm on a side, consisting of two identical, 2550 $\times$ 1275~mm, sliding half-doors. The half-doors slide apart in opposite directions, giving access to the SSC. Along the periphery of the half-doors, lead radiation protection bricks, 200~mm high, are installed. Inside them, within the square space 2450~mm on the side, three layers of shielding, one above the other, are placed: 160~mm of borated polyethylene, 50~mm of lead and 40~mm of pure polyethylene. The upper part of the passive shielding is mounted separately in each half-door. The total weight of shielding materials in each half-door is 2.7 tons. Their sliding is assured by an electrical drive.

\subsection{Muon veto}
Sidewise cosmic muons are detected in iDREAM by the GC. Meanwhile, for straight vertical muons that directly cross the buffer and then the TG, the detector is equipped with two scintillation polymethylmethacrylate plates. Each plate, 1900 $\times$ 1200 $\times$ 33~mm in size, is placed on the top of each half-door above the shielding. The plates are sequentially wrapped with a reflective Mylar tape, light-protective black polyethylene film and a synthetic leather. On the two side edges, the plates are enclosed with an aluminum support, in which six PMTs (PMT-85), three on the side, are installed. To provide an optical contact between the photocathodes and the edges of the plate, a thin layer of optical grease based on quartz vaseline is used.\par
\begin{figure}[htbp]
\centering 
\includegraphics[scale=0.35]{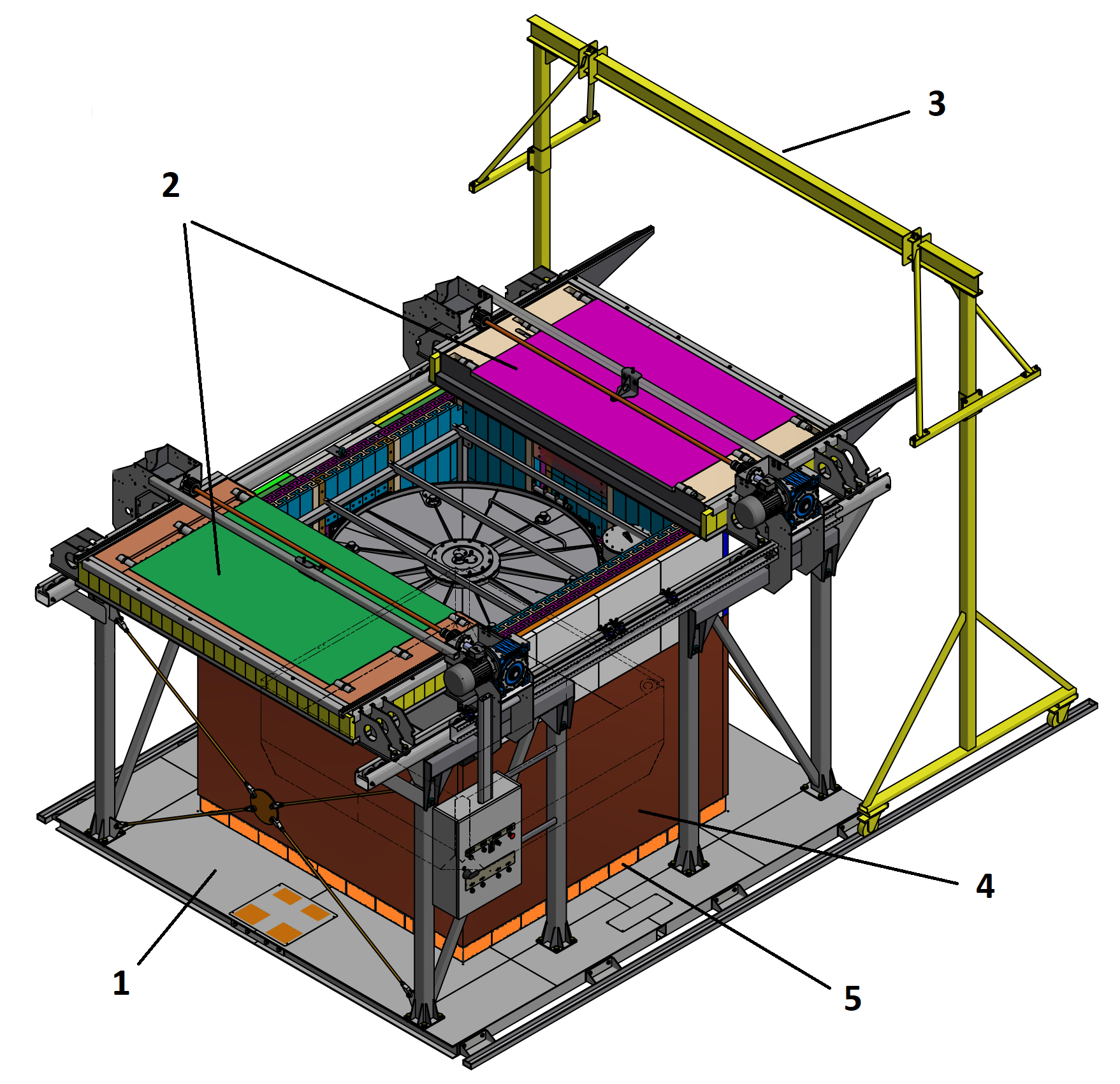} 
\caption{\label{fig4}Sketch of the detector inside the shielding: 1 -- skid platform, 2 -- muon plates on the top of the sliding half-doors, 3 -- gantry mobile crane, 4 -- pure polyethylene (outermost shielding layer), 5 -- cast iron shielding layer.}
\end{figure}

Figure \ref{fig4} shows the detector inside the full shielding. Two sliding half-doors with the muon veto plates installed above are shown opened and the detector is seen inside. The gantry mobile crane used to mount the detector is seen on the right.\par

\section{Liquid scintillator}
The scintillator used as a neutrino target in the iDREAM detector was prepared on the basis of linear alkylbenzene (LAB), gadolinium (3,5,5-trimethyl)hexanoate and fluorescent compounds, 2,5-Diphenyloxazole (PPO) and 1,4-Bis(2-methylstyryl)benzene (bis-MSB).

LAB is a mixture of hydrocarbons produced by the reaction of benzene and alkenes in the presence of catalysts. The number of individual hydrocarbons that make up the technical LAB is more than 40. The chemical structure of these compounds is due to the nature of the alkenes used for the alkylation of benzene, which are unsaturated hydrocarbons with a terminal unsaturated bond (mainly) with the number of carbon atoms from 6 to 20. Thus, the process of obtaining LAB can be represented as follows: C$_{6}$H$_{6}$ + CH$_{2}$ = CH-R = C$_{6}$H$_{5}$-CH (CH$_{3}$) R, where R is a saturated unbranched hydrocarbon radical with 4 to 16 carbon atoms, mainly from 10 to 14.

In order to prepare a concentrated solution of gadolinium (3,5,5-trimethyl)hexanoate, the technical LAB was purified by vacuum distilling. Gadolinium (3,5,5-trimethyl)hexanoate was prepared starting from gadolinium oxide containing 99.999\% Gd and (3,5,5-trimethyl)hexanoic acid in a multistep process.

Initially, an aqueous solution of gadolinium chloride was obtained. For this, high-purity gadolinium oxide was treated with heating and vigorous stirring with hydrochloric acid (high purity grade). An excess of gadolinium oxide was used to rule out an excess of hydrogen chloride dissolved in the solution. The resulting gadolinium chloride solution was reacted with an aqueous solution of the ammonium salt of (3,5,5-trimethyl)hexanoic acid.

A solution of the ammonium salt of (3,5,5-trimethyl)hexanoic acid was obtained by neutralizing (3,5,5-trimethyl)hexanoic acid with an aqueous solution of ammonia (high purity grade) in the distilled water. Prior to use, the (3,5,5-trimethyl)hexanoic acid was purified by vacuum distillation.

The interaction of aqueous solutions of gadolinium chloride and ammonium (3,5,5-trimethyl) hexanoate yielded gadolinium (3,5,5-trimethyl)hexanoate insoluble in water. This product was filtered, washed several times with distilled water and then dried under vacuum over potassium hydroxide, which was used as a desiccant. The resulting product was dissolved in tetrahydrofuran. In this case, some of the impurities, including the decomposition products of gadolinium (3,5,5-trimethyl)hexanoate, formed during the dehydration process, as well as impurities not removed during the washing process of the crude gadolinium (3,5,5-trimethyl)hexanoate, were separated by filtration of the solution, since these compounds are insoluble in tetrahydrofuran.

The resulting solution of dry gadolinium (3,5,5-trimethyl)hexanoate in tetrahydrofuran was mixed with LAB purified by vacuum distillation. Then, the tetrahydrofuran was removed by evaporation under reduced pressure. Thus, a master solution of gadolinium (3,5,5-trimethyl) hexanoate in LAB was obtained, containing 10 g/l of Gd (per element).

A concentrated solution of gadolinium (3,5,5-trimethyl)hexanoate was used to prepare a scintillation solution by mixing with a solution of PPO and bis-MSB in LAB. The resulted iDREAM scintillator in the TG contains 1 g/l of Gd, 2.7 g/l of PPO and 0.02 g/l of bis-MSB.

\section{Detector subsystems}
\subsection{Detector filling and nitrogen purging systems}
Detector filling is performed using dual diaphragm pumps connected to the FDH. The FDH's three-way valves are connected via bellow pipes to the stainless steel tubes, each going up to the respective detector volume: the TG, the GC and the buffer. When the detector volumes are filled/emptied, the valves are switched so that the tubes are connected to the pumps. During the data taking, the valves connect the tubes to the free volume under the SSC, thereby equalizing the pressure of the fluids. To depressurize the detector, one can switch a dedicated valve to atmospheric volume.

The detector volumes are filled using a flow meter which has an accuracy of $\pm$ 0.5\%. After filling, the TG and GC are purged with pure nitrogen to remove oxygen from the scintillator, which may affect its chemical stability. The nitrogen is supplied by circular separators, the ring coiled stainless steel tubes with holes evenly distributed along its length. The separators are installed on the floor of the IT and the GC, and connected to the FDH through vertical stainless steel tubes with bellow pipes at their upper ends. The nitrogen is flowed through the TG and the GC at a pressure of about 10 kPa.

To protect the scintillator from outside air, the volume under the SSC is kept in the nitrogen atmosphere within the range of 0.5$\div$1.5 kPa by means of the solenoid valve. The nitrogen bottle is permanently connected to the FDH through the valve which automatically opens at the bottom threshold and closes at the top threshold. The bottom and top valve activation thresholds are set by the operator in the interface window of the slow control software (see Section~5.2). The current overpressure values are displayed on the screen and archived into a file.

The filling of the detector at Kalinin NPP was performed in the following way. First, both the TG and GC were filled with the gadolinium-free scintillator, whereas the buffer was filled with pure LAB. At this point, the detector hermeticity was tested and preliminary data were taken. After hermeticity was guaranteed, a part of the scintillator was removed from the TG and replaced by a concentrated solution of gadolinium (3,5,5-trimethyl)hexanoate in LAB. The newly prepared Gd-LS was then flushed with nitrogen.

\subsection{Slow control system}
The slow control (SC) system is designed to operate and control readings from all detector sensors (e.g. level and pressure sensors) in a single detector-to-computer data exchange protocol. The SC system is realized via Controller Area Network (CAN) bus. It is a bus type network in which all nodes can transmit and receive data. An open protocol CANopen, easily ported to required devices, is used as a top level protocol.

The SC system includes a CAN controller board with a standard PCI I/O bus for connecting peripheral devices to the computer motherboard, and a set of microcontrollers linking all detector sensors and electronic control module (ECM) to the CAN controller. The devices used are as follows:

\begin{itemize}
  \item level and temperature sensor in the TG;
  \item level sensors in the GC, buffer, and emergency tray;
  \item nitrogen pressure sensor under the SSC;
  \item digital pressure gauge on the nitrogen bottle;
  \item solenoid valve connecting the FDH to a nitrogen bottle;
  \item temperature sensor in the experimental hall;
  \item atmospheric pressure sensor in the experimental hall.
\end{itemize}

The SC system software interface is written in LabView. It is linked to the ECM and allows to:

\begin{itemize}
  \item control the detector pumps when filling/emptying the detector;
  \item control the solenoid valve when filling the volume under the SSC with nitrogen;
  \item control the technical parameters of the detector, their treatment and the display of the data in real time.
\end{itemize}

The SC software allows monitoring of changes in sensor readings in offline mode. Data are read and saved in a file with a tunable time interval. 

\subsection{Data acquisition system}
The iDREAM data acquisition (DAQ) system is a distributed network system for data taking and primary analysis. It consists of the following subsystems:
\begin{itemize}
  \item PMTs high-voltage power supply (HVPS) subsystem;
  \item subsystem for primary processing of the PMT signals;
  \item trigger subsystem;
  \item waveform digitizer (WFD).
\end{itemize}

\begin{figure}[htbp]
\centering 
\includegraphics[scale=0.35]{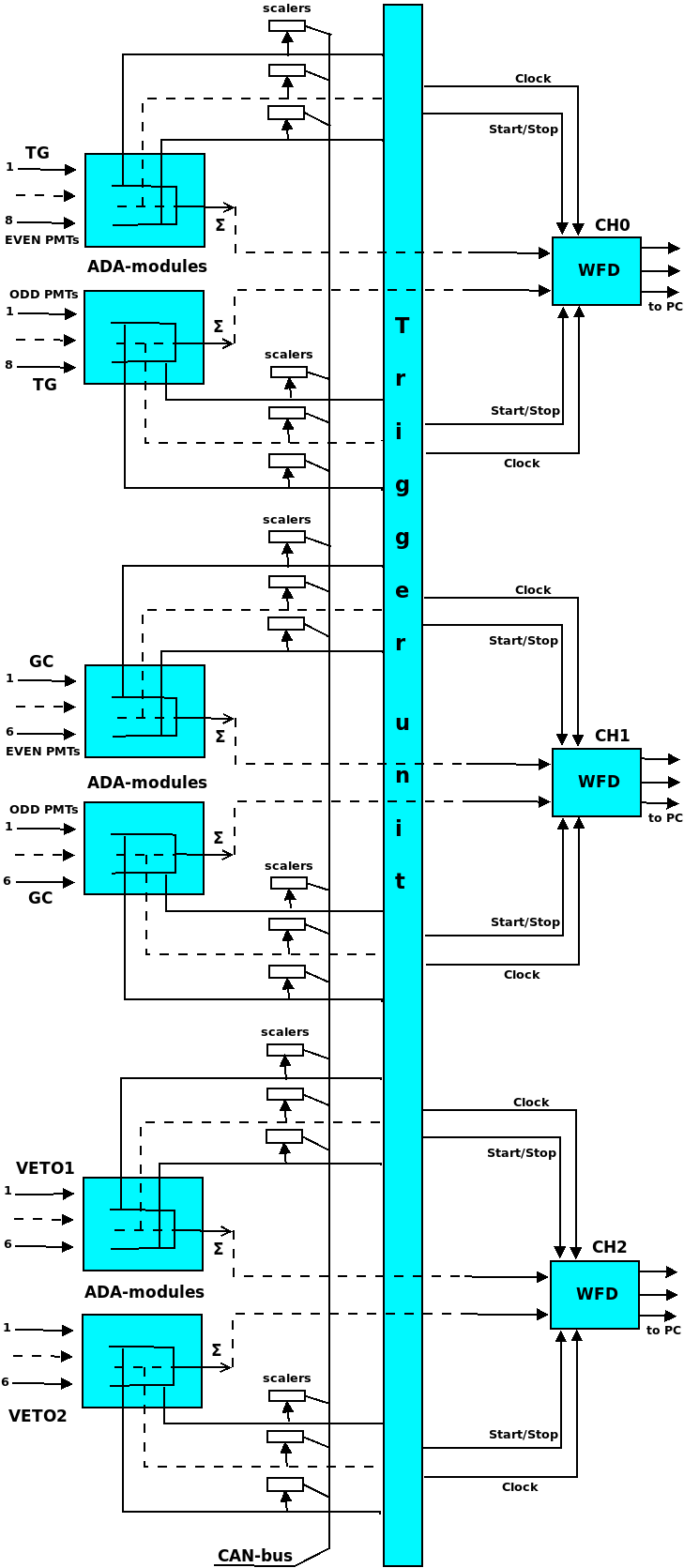} 
\caption{\label{fig5} DAQ system structural diagram.}
\end{figure}

DAQ system structural diagram is shown in figure \ref{fig5}. The PMT signals of the TG, GC and muon veto go to the amplifier-discriminator and adder (ADA) modules. The logical LVDS signals activated in the ADA module when the PMT discriminators are triggered, go to the trigger unit (TU) where the trigger scheme is formed. If a trigger signal has been generated, the summed PMT signals outputted from the ADA modules to the WFD are digitized and stored in the PC memory.

The DAQ system is controlled by the dedicated software using a set of web interface commands, which interact with different detector processes. The whole web interface runs on the Linux Debian 8.11 operating system. The amount of data transmitted does not exceed 100 Mbit/s, which matches the capacity of the Kalinin NPP PC network. 

Below, the iDREAM DAQ subsystems are described.
    
\subsubsection{HVPS subsystem}
The PMTs are connected to the DAQ according to an individual schematic, in which the HV power supply of the PMTs and their signals are made on a single cable. The HV decoupling nodes are installed in the ADA modules (see Section~5.3.2).

The HVPS subsystem is based on a modular design. It consists of six 8-channel 1U modules manufactured by Marathon LLC (Russia). Each module provides a 200 to 2000 V supply voltage range with a DC current consumption of up to 1.0 mA. Output voltage adjustment accuracy is 1 V and stabilization accuracy is 0.2\%. The operating voltage range for the PMT is between 1500 V and 1800 V. The number of HVPS channels corresponds to the number of PMTs plus redundant channels (48 channels in total, including 16 channels for the TG PMTs, 12 channels for the GC PMTs, and 12 channels for the muon plates PMTs). Voltage dividers are manufactured on printed circuit boards, and their current consumption at 2000 V is 0.5 mA. Each HV channel is individually regulated and monitored with an accuracy of 0.2\% of the current and voltage measurement.

In order to control the HVPS, the CAN-bus serial bus connected to the CAN-PCI controller with the top level protocol CAN-open and dedicated software installed on the PC is used. The software allows to switch on and off any individual channel, set the required voltage and keep it in memory, monitor the output voltage, the current consumed by the PMTs, and the temperature in the module.

\subsubsection{Subsystem for primary processing of the PMT signals}
\begin{figure}[htbp]
\centering 
\includegraphics[scale=0.4]{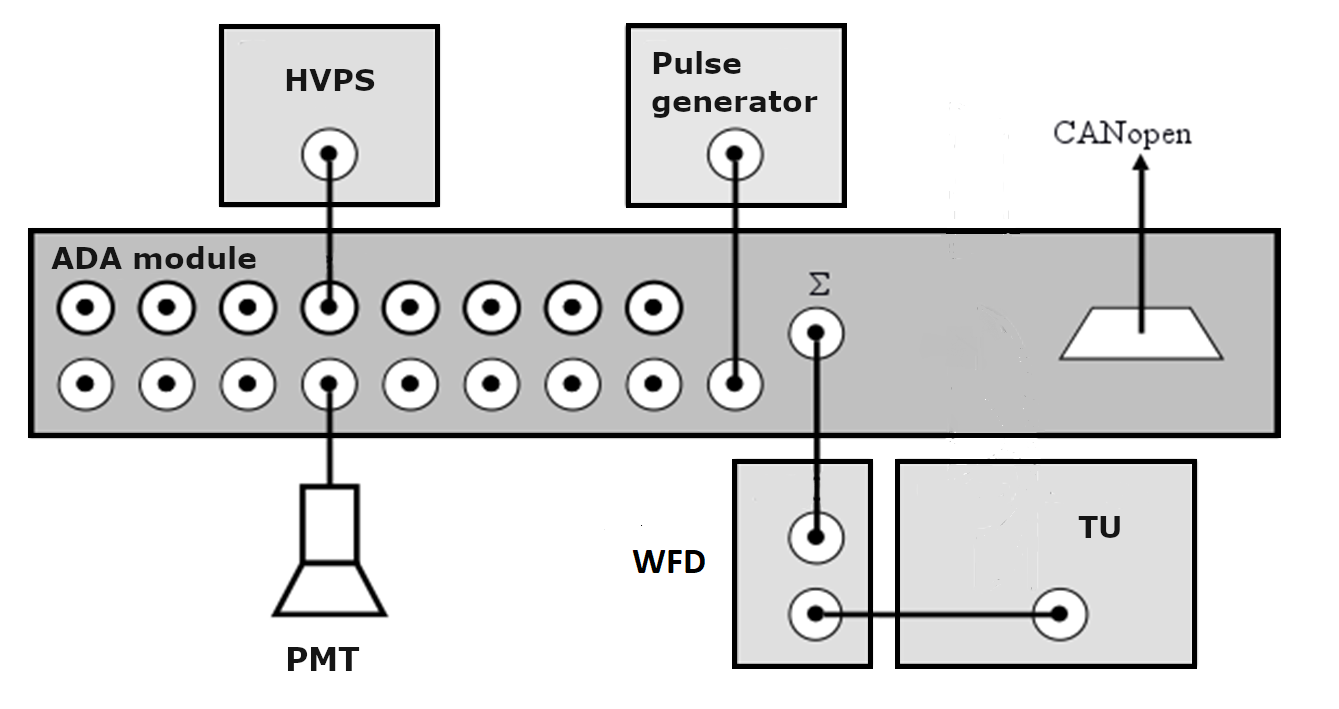} 
\caption{\label{fig6} Schematic drawing for connecting a PMT and HVPS through the ADA module.}
\end{figure}

The primary processing system for the PMT signals is constructed in a modular way. It consists of five 8-channel 2U ADA modules, manufactured by Marathon LLC (Russia) specially for iDREAM. A schematic drawing of the connection of the PMT and HVPS through the ADA module is shown in figure \ref{fig6}. 

The ADA module performs the following operations:
\begin{itemize}
  \item decoupling of the HV and signal from the PMTs;
  \item amplification, branching and analog summation of the input analog signals from the PMTs;
  \item discrimination of the PMT signals with an adjustable threshold;
  \item logic pulses rate measurement in each PMT channel.
\end{itemize}

The module consists of eight signal amplifiers, whose outputs are used to generate logical signals. One more amplifier sums the signals of all eight analog channels. In addition, the module contains eight logical outputs of adjustable threshold discriminators, used for adjusting and monitoring the PMTs state. Eight module inputs are connected to the HVPS modules, while the outputs are connected to the PMTs.

The module can operate either in manual mode or in software-controlled mode. In manual mode, the discrimination thresholds are set manually by adjusting eight potentiometers. The indication allows to observe the states of the HV inputs and the discriminator outputs. In software-controlled mode, the ADA module represents a CAN-open node. In this mode one is able to:
\begin{itemize}
  \item set the individual discrimination levels and the offset at the input of the sum;
  \item measure the pulse repetition rate in accordance with the discrimination thresholds;
  \item monitor HV on/off state at the corresponding inputs;
  \item control the temperature of the power supply.
\end{itemize}

The outputs of the discriminators are fed to the counters of the PMT logical signals. The PMT count rate data are transmitted from each module's channel to the SC system via the CAN-bus network.

\subsubsection{Trigger subsystem}
The TU was developed specifically for the iDREAM project. It is based on a modern approach using field programmable gate array (FPGA), which allows to distinguish rare useful events with a complex signature from the general data flow. The TU is based on the Xilinx Zynq-7020 chip. The output trigger signal is generated based on a complex configurable logic that takes into account the signals from the discriminators and other detector systems, a total up to 97 input and output condition signals.

Sources for preliminary triggers may include signals from the TG PMTs, the GC PMTs, the muon veto PMTs and/or their combinations. If necessary, additional auxiliary detectors can be connected to the TU. A unique identifier (trigger type) is assigned to each pre-trigger source. Thus, for each detected event, the trigger sources or their combination is known.

The preliminary trigger is activated if the number of triggered discriminators exceeds the level specified by the user (the majority level). The final trigger is formed from the preliminary triggers depending on the type of measurements performed. 

Such an approach to trigger generation offers maximum functionality, scalability, and adjustment flexibility. The ARM architecture processor core included in the Zynq-7020 chip is used to control and monitor the TU, connecting it to other components of the DAQ system via Ethernet. The peculiarity of the developed trigger system is a very wide range of TU parameter configuration, defined through the device's configuration registers without reprogramming the FPGA. All key trigger parameters such as signal delays, trigger gate length, majority threshold level, coincidence-anticoincidence circuit parameters etc. can be set by the operator. By varying the parameters, one can quickly change the class of events registered by the detector, if necessary. This is of particular importance during initial fine tuning and detector calibration. The combination of an FPGA based trigger and fast parallel digitizers has created a high performance DAQ system with no dead time. 

The trigger signals are received at the control input of the WFD, which leads to the data being read into the memory of the control computer.

\subsubsection{WFD}
The WFD module used is 8 channels 14 bits 500 MHz CAEN DT5730. The events are digitized in a 600 ns DAQ gate with the trigger position located around 180 ns from the beginning of the gate. The data are transferred to the control computer by optical link, providing up to 80 MB/s transfer rate. This guarantees no loss of data at iDREAM trigger rate below $\sim$100 kHz.

Two WFD channels digitize the analog sum signals of the TG and GC PMTs separately, while the third channel digitizes the PMT sum signal of the two muon plates. Therefore, each recorded event consists of a set of digitized data from the detector. This allows offline analysis of any different combination of measured signals to better recognize the IBD event.

\subsection{Calibration system}
In order to calibrate the detector and monitor its response, a special calibration system has been developed. By default, the system is installed in the TG calibration channel, but can easily be moved to any GC calibration channel. The system allows to determine the position of the radioactive source along the {\it z} axis with an accuracy of 2 mm.

\begin{figure}[htbp]
\centering 
\includegraphics[scale=0.62]{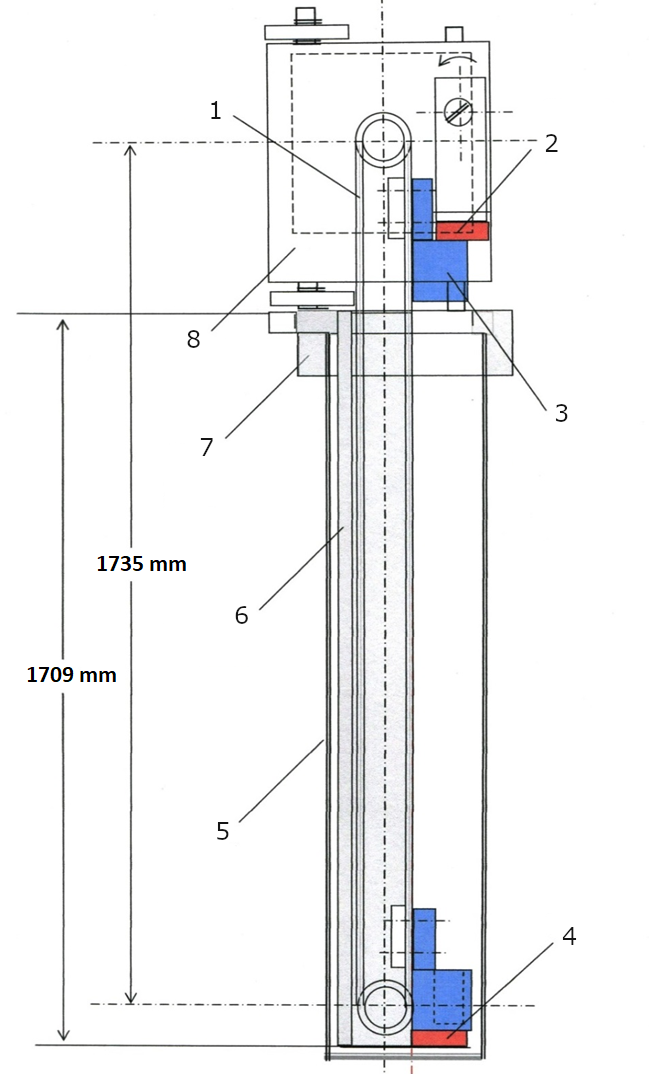} 
\caption{\label{fig7}Schematic drawing of the iDREAM calibration system: 1 -- toothed belt, 2 -- top seal switch, 3 -- shuttle with radioactive source, 4 -- bottom seal switch, 5 -- dry stainless steel channel immersed in the TG or GC, 6 -- extension of the stepper motor holder, 7 -- stepper motor holder, 8 -- flange of the stepper motor.}
\end{figure}

The calibration source encapsulated in a shuttle is mounted on a closed toothed belt enfolding two pulleys. The top pulley is located on the shaft of the stepper motor, while the bottom is on the extension of the motor holder at the bottom part of the calibration channel, as shown in figure \ref{fig7}. In its uppermost and lowermost positions, the shuttle turns on the seal switch and stops the stepper motor. The motor power supply operates at 24 V DC.

The calibration system is controlled from the operator panel with the dedicated software installed. In a separate software window, the date and time of the calibration, the calibration mode and the position of the source are shown. The software provides two calibration modes. In both modes, the interface displays the calibration channel scale in mm and the current position of the shuttle with the source at the scale.

In mode 1, calibration is performed at any chosen point along the vertical axis of the calibration channel. In this mode, the operator presets the calibration point. The position of the set point is calculated at the height starting from the lowermost position of the shuttle.

In mode 2, the scan of the detector in the selected range with a fixed step is performed. In this mode, the operator can set the scan step in the range of 50 to 150 mm. The maximum vertical length of the calibration range is 1200 mm above the lowest position of the shuttle, i.e. inside the membrane tube.

\section{Detector calibration}
The detector design provides the TG calibration along the vertical axis ({\it x} = {\it y} = 0). We investigated the detector response to the $\gamma$-ray calibration sources $^{137}$Cs ($E_\gamma$ = 0.662 MeV), $^{54}$Mn ($E_\gamma$ = 0.835 MeV), $^{65}$Zn ($E_\gamma$ = 1.115 MeV), and $^{60}$Co ($E_\gamma$ = 1.173, 1.332 MeV), as well as $^{252}$Cf fast neutron source. The sources were specially manufactured for iDREAM in cylindrical containers 7~mm in diameter and 10~mm in height, in order to fit with the shuttle of the calibration system. 

\begin{figure}[htbp]
\centering 
\includegraphics[scale=0.22]{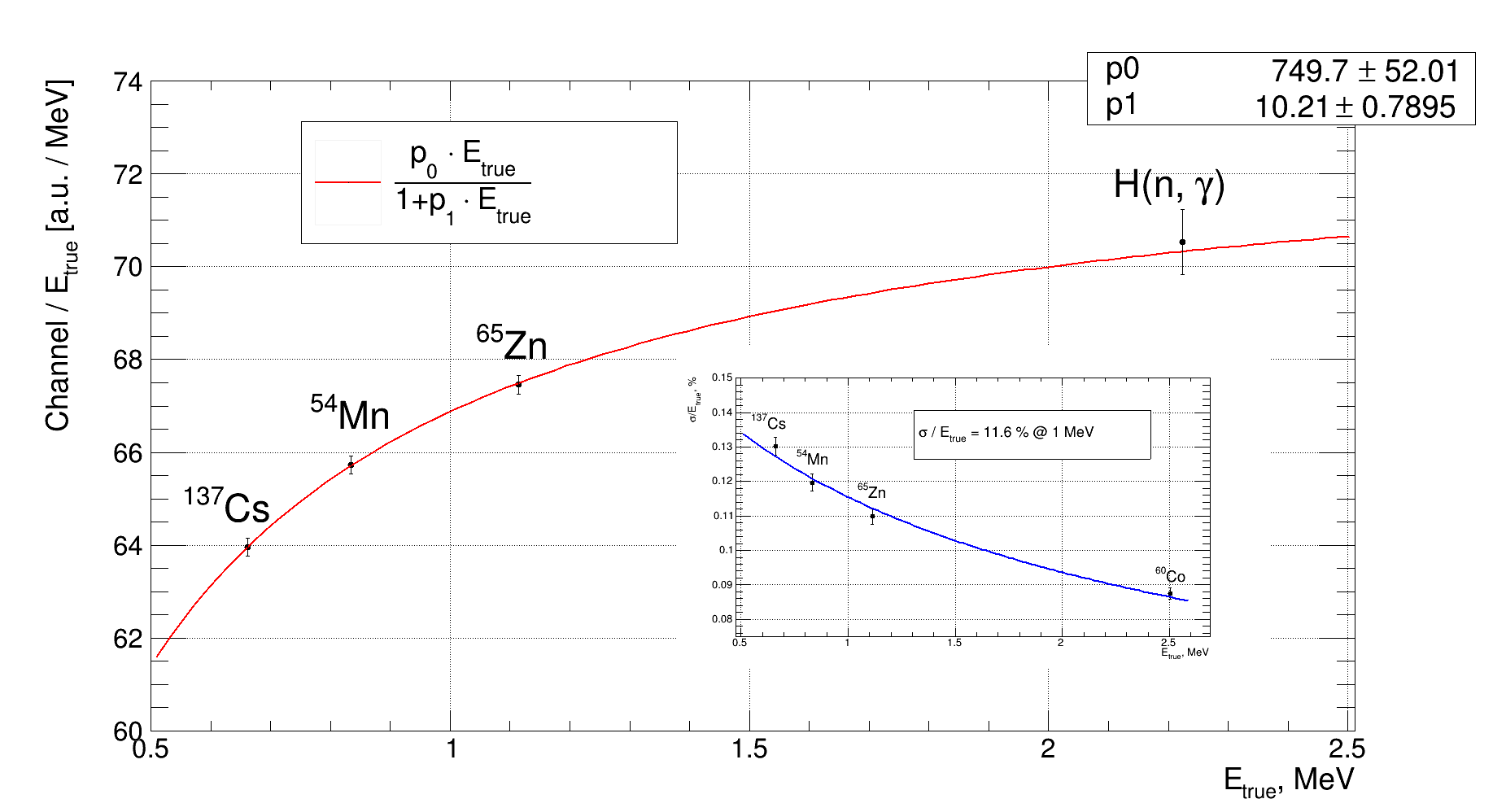} 
\caption{\label{fig8}Detector response function determined from 4 data points of $^{137}$Cs, $^{54}$Mn, $^{65}$Zn, and {\it n}H capture $\gamma$-lines. The inset shows the observed detector resolution as function of $E_\gamma$.}
\end{figure}

The non-linearity of the iDREAM energy scale was defined as the dependence of the ratio $E_{measured}/E_{true}$ on $E_{true}$, measured in the detector center. The response function was determined from 4 data points of $^{137}$Cs, $^{54}$Mn, $^{65}$Zn, and {\it n}H capture $\gamma$-lines. The latter point ($E_\gamma$=2.224 MeV) was obtained from the calibration with the $^{252}$Cf neutron source. As shown in figure \ref{fig8}, the data points are well described using simple "Birks-like"~\cite{Birks} function $f(E_{\gamma}) = p_{0} \cdot E_{\gamma}/(1 + p_{1} \cdot E_{\gamma})$. The inset shows the observed detector resolution as function of $E_\gamma$. The obtained resolution at $E$ = 1 MeV equals to $\sigma/E$ = 11.6\%.

\begin{figure}[htbp]
\centering 
\includegraphics[scale=0.22]{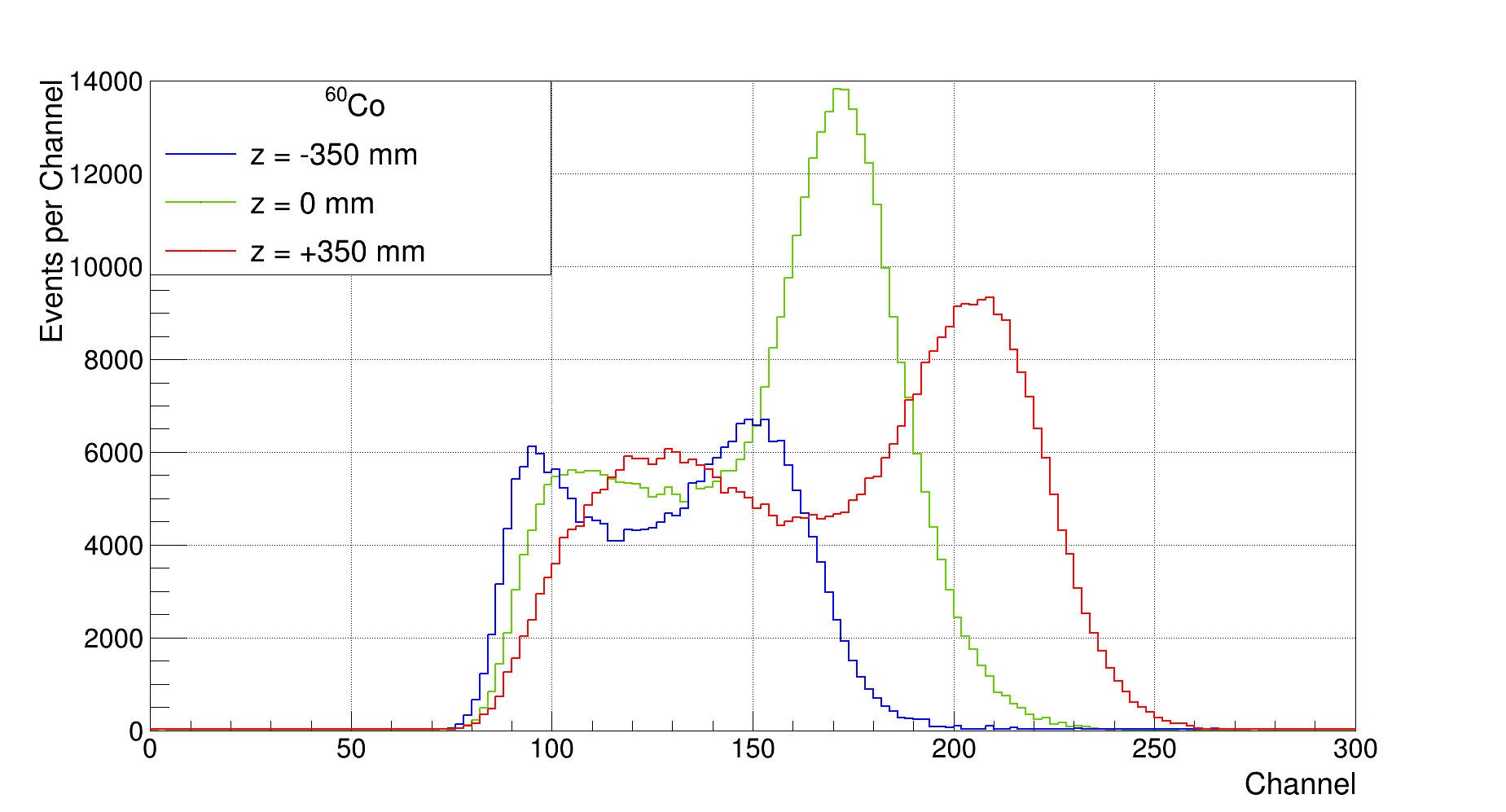} 
\caption{\label{fig9}Response of the detector to the $^{60}$Co $\gamma$-ray source deployed in {\it z} = -350, 0, +350 ~mm ({\it x} = {\it y} = 0).}
\end{figure}

Figure \ref{fig9} shows the detector response to the $^{60}$Co source deployed at the bottom ({\it z} = -350 ~mm, blue curve), in the center ({\it z} = 0 ~mm, green curve), and at the top of the TG below the membrane ({\it z} = +350 ~mm, red curve). In all cases, the rightmost peak corresponds to the total absorption of the both 1.173 and 1.332 MeV $\gamma$-quanta. Since PMTs only see the TG from one side, the geometric heterogeneity of the light collection was expected. Indeed, for the cylindrical part of the TG enclosed within {\it z} = (-350, +350)~mm the observed heterogeneity of the light collection was found to be (-12, +20)~\%.

\begin{figure}[htbp]
\centering 
\includegraphics[scale=0.22]{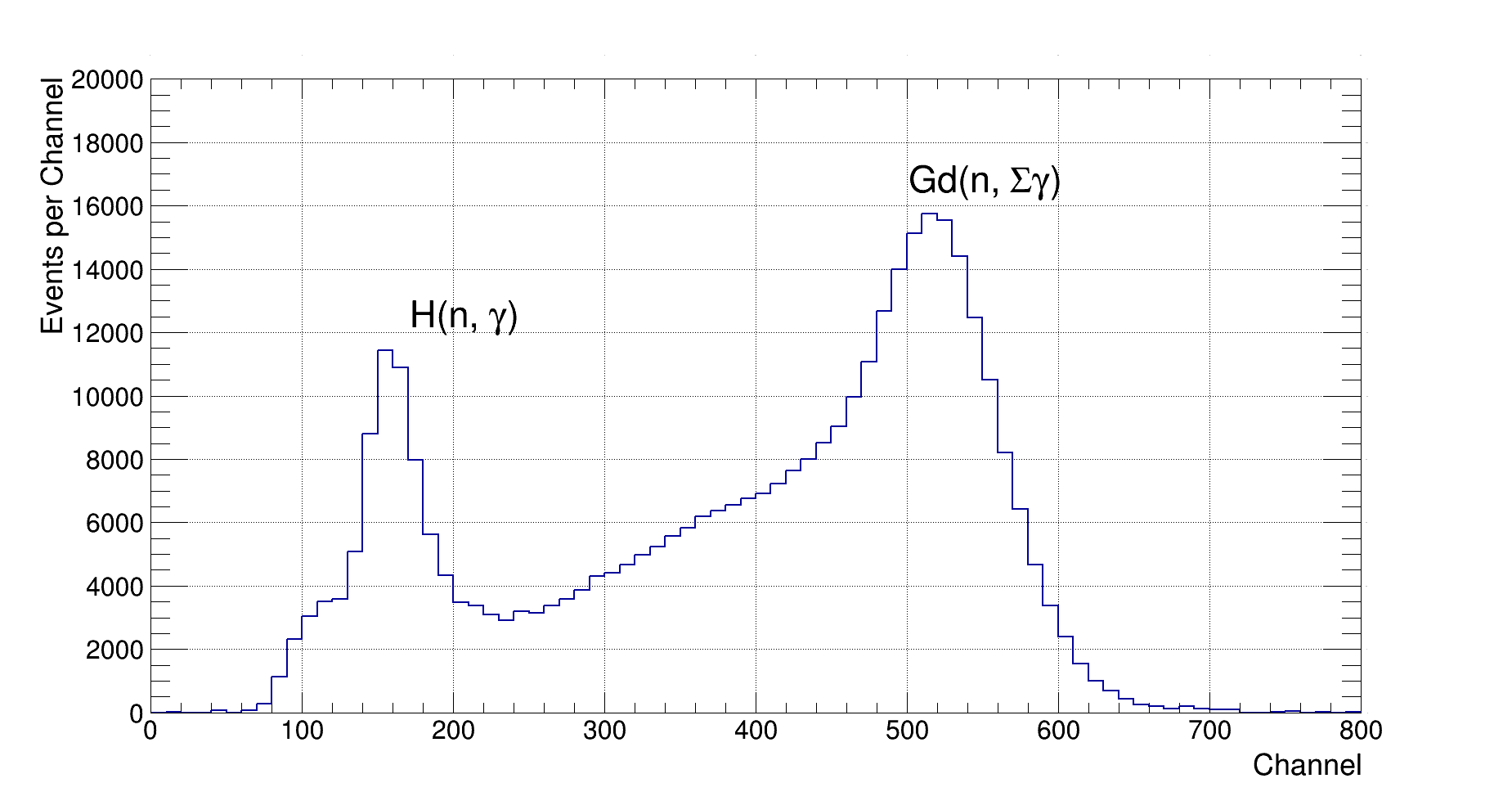} 
\caption{\label{fig10}Response of the detector to the $^{252}$Cf fast neutron source deployed in the center of the TG.}
\end{figure}

\begin{figure}[htbp]
\centering 
\includegraphics[scale=0.22]{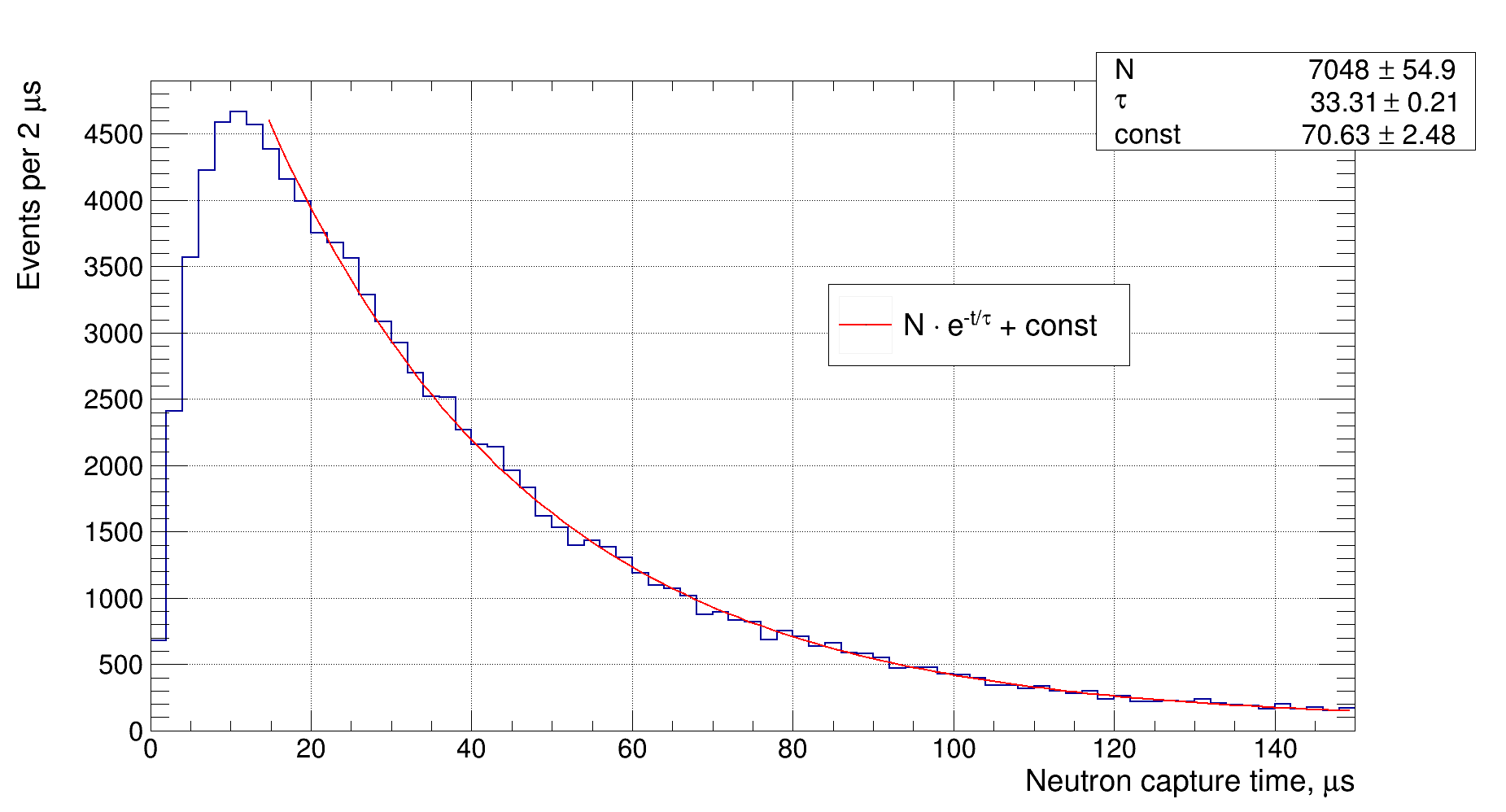} 
\caption{\label{fig11}Neutron capture time spectrum of the $^{252}$Cf source deployed in the center of the TG.}
\end{figure}

Further investigations were performed with the $^{252}$Cf fast neutron source. Because neutron capture time defines the length of the time window between prompt and delayed IBD events, it is of crucial importance for the antineutrino analysis. The $^{252}$Cf neutron capture candidate events were selected in a time window of [1$\div$150] $\mu$s following prompt spontaneous gamma rays that accompany $^{252}$Cf fission. The detector response to the $^{252}$Cf source deployed in the center of the TG is shown in figure \ref{fig10}. Two peaks corresponding to the neutron captures on H and Gd are clearly observed. The neutron capture time spectrum is shown in figure \ref{fig11}. Fitting the spectrum with $f(t) = N \cdot e^{-t/\tau} + const$ gives for the measured neutron capture lifetime $\tau$ = 33.3$\pm$0.2 $\mu$s, in concordance with the one expected for the Gd concentration of 1 g/l in the TG. We found no change in the neutron capture lifetime relative to the source position within the calibration channel.

\section{Conclusion}
In summary, the description of the iDREAM detector and its systems, many of which were specifically designed for the project, was presented. To ensure easy maintenance of the detector at an NPP, simple design and technologies have been used throughout the whole iDREAM facility. The detector calibration was performed within the vertical calibration channel inside the TG, as provided by the detector design. The obtained resolution at $E$ = 1 MeV equals to $\sigma/E$ = 11.6\%, while the heterogeneity of the light collection for the cylindrical part of the TG enclosed within {\it z} = (-350, +350)~mm was found to be (-12, +20)~\%.

We note that iDREAM was created with the objective of remote monitoring of power reactors. It is currently the one of the few very short baseline neutrino experiments ongoing at a power plant. Studies with this prototype detector will be an important step in the development of commercial neutrino-based devices for the nuclear industry.


\acknowledgments
The iDREAM experiment at Kalinin NPP is supported by the Russian Science Foundation (project No. 22-12-00219). Data analysis is performed using computational resources of MCC NRC "Kurchatov Institute", http://computing.nrcki.ru/. The authors are grateful to ROSATOM State Corporation and Kalinin NPP staff for their support in setting up the iDREAM installation and taking measurements.



\end{document}